\documentclass[prl,twocolumn,aps,a4paper,superscriptaddress,showpacs]{revtex4}
\usepackage{amsmath}
\usepackage[dvips]{graphicx}
\begin{document}

\newcommand\beq{\begin{equation}}
\newcommand\eeq{\end{equation}}
\newcommand\bea{\begin{eqnarray}}
\newcommand\eea{\end{eqnarray}}
\newcommand\bseq{\begin{subequations}} 
\newcommand\eseq{\end{subequations}}
\newcommand\bcas{\begin{cases}}
\newcommand\ecas{\end{cases}}

\title{Dark Matter Prediction from Canonical Quantum Gravity with Frame Fixing}

\author{Giovanni Corvino}
\author{Giovanni Montani}
\email{montani@icra.it} 
\affiliation{Dipartimento di Fisica Universit\`a di Roma ``La Sapienza''}
\affiliation{ICRA---International Center for Relativistic Astrophysics  
c/o Dipartimento di Fisica (G9) Universit\`a di Roma ``La Sapienza'',
Piazza A.Moro 5 00185 Rome, Italy}

\today

\begin{abstract}

We show how, in canonical quantum cosmology, the frame fixing induces a new energy density contribution having features compatible with the (actual) cold dark matter component of the Universe. First we quantize the closed Friedmann-Robertson-Walker (FRW) model in a sinchronous reference and determine the spectrum of the super-Hamiltonian in the presence of ultra-relativistic matter and a perfect gas contribution. Then we include in this model small inhomogeneous (spherical) perturbations in the spirit of the Lemaitre-Tolman cosmology. The main issue of our analysis consists in outlining that, in the classical limit, the non-zero eigenvalue of the super-Hamiltonian can make account for the present value of the dark matter critical parameter. Furthermore we obtain a direct correlation between the inhomogeneities in our dark matter candidate and those one appearing in the ultra-relativistic matter.

\end{abstract}



\maketitle

\section{Basic Statements}

In recent works \cite{Mon2002,CosmIssues,ProcGross,Dualism,Korean,MM04} it was outlined the relation existing in quantum gravity between the frame fixing and the appearence of an evolutive Schr\"odinger dinamics for the wave functional. In particular in \cite{ProcGross,Dualism} it has been stressed how, explicitly breaking the time displacements invariance of theory, corresponds to introduce a reference fluid in the dynamics. Within such a framework, the quantum evolution of a 3-geometry $\{h_{ij}\}$ is provided by the following equation for the wave functional $\psi(t,\{h_{ij}\})$

\begin{equation}
i\hbar\partial_{t}\psi=\int_{\Sigma_{t}^{3}}d^{3}x {N\hat{H}},
\end{equation}

where a compact 3-hypersurface $\Sigma_{t}$ is associated to any value of the label time $t$ and by $\hat{H}$ and $N$ we denoted, respectively, the super-Hamiltonian operator and tha lapse function; since $\psi$ depends only on 3-geometries, then it is anihilated by the super-momentum operator, i.e. $\hat{H}_{i}\psi=0$. If we take the expansion

\begin{align}
\psi\left(  t,\{h_{ij}\}\right)  =\underset{y}{%
{\displaystyle\int}
}D\Omega &  \Theta\left(  \Omega\right)  \chi\left(\Omega,
\{h_{ij}\}\right)  \cdot\nonumber\\
&  \cdot\exp\left\{  -\frac{i}{\hbar}\int\limits_{t_{0}}^{t}dt^{\prime
}\underset{\Sigma_{t}^{3}}{\int}d^{3}x\left(  N\Omega\right)
\right\}\,  ,\label{espansione}%
\end{align}

being $D\Omega$ the Lebesgue measure on the functional space $\mathcal{Y}$, then we get an eigenvalue problem of the form

\begin{equation}
\hat{H}\chi=\Omega(x^{i})\chi,\quad \hat{H}_{i}\chi=0. \label{autovalori0}
\end{equation}

As shown in \cite{Dualism}, the presence of non-zero eigenvalues for the super-Hamiltonian operator, can be restated, in tha classical limit, in terms of Einstein equation with a source. More precisely it outcomes a dust fluid whose energy density reads as

\begin{equation}
\varepsilon\equiv -\frac{\Omega(x^{i})}{2K\sqrt{h}},
\end{equation}

where $K$ denotes the Einstein constant. The non relativistic nature of such energy density and its quantum origin lead (see \cite{CosmIssues}) to candidate it as a possible explanation for the Universe dark matter component.

In this work we show how it is possible to have a closed FRW model for which the classical limit of the super-Hamiltonian eigenvalue has the appropriate features to account of the cold dark matter (actual) energy density. In section 2 we quantize the closed FRW model in agreement to the equations (\ref{autovalori0}) and including in the quantum dinamics two phenomenological terms, one accounting for the ultra-relativistic matter and another one corresponding to a perfect gas component. The former term describes the thermal bath of fundamental particles, while the latter is included because the quantum behavior of the Universe is expected in the Planck era and then a non-relativistic Planck gas (i.e. a perfect gas of particles with the Planck mass) is reliably postulated. In section 3 we investigate the behavior of small (spherical) perturbations, by the framework of a Lemaitre-Tolman cosmology. The inhomogeneous corrections to the super-Hamiltonian operator is calculated via a perturbation theory of the eigenvalues problem. In section 4 we show how the actual value of the Universe dark matter energy density \cite{RobRot1973,OstPee1974,EinKaa1974,RubTho1978} can be fitted by our analysis.

A relevant issue of the perturbation theory is shown to be the direct correlation between the inhomogeneities of the ultra-relativistic matter and those ones emerging in the dust fluid energy density, i.e. our candidate for dark matter.

\section{Canonical Quantization of the Closed FRW Model}

In this section we will quantize the closed FRW model into a sinchronous reference and in the presence of ultra-relativistic matter and a perfect gas contributions. The line element of such a model takes the form:

\begin{equation}
ds^{2}=-N^{2}dt^{2}+R_{c}^{2}(t)[d\xi^{2}+\sin^{2}\xi(d\eta^{2}+\sin^{2}\eta
d\phi^{2})]\,,\label{yyy}%
\end{equation}

being $N$ the lapse function, $R_{c}$ the Universe radius of curvature and with $0<\xi<\pi$, $0<\eta<\pi$, $0<\phi<2\pi$. The dinamics associated to this line element is summarized by the action:

\begin{equation}
S=\int_{\Sigma_{t}^{3}\times R} dtd^{3}x\left\{p_{R_{c}}\frac{\partial R_{c}}{\partial t}-NH\right\}\, ,\label{azione}
\end{equation}

where the super-hamiltonian reads

\begin{equation}
H=-\frac{l_{Pl}^{2}}{3\pi\hbar}\frac{p_{R_{c}}^{2}}{R_{c}%
}+\frac{\mu^{2}c}{R_{c}}-\frac{3\pi\hbar}{4l_{Pl}^{2}}R_{c}+\frac{\sigma^{2}c}{R_{c}^{2}}\,,\label{w}%
\end{equation}

where $\mu^{2}, \sigma^{2}=const.$ Here we introduced the term $\mu^{2}c/R_{c}$ in order to make account phenomenologically of the primordial termal bath, which in the Planck era is well described by an ultra-relativistic fluid. Furthermore we add to the dinamics a perfect gas contribution in order to represent a gas of Planck mass particles which in the Planck era have non-relativistic properties. For the FRW model the eigenvalues problem (\ref{autovalori0}) reads as

\begin{equation}
\hat{H}\chi(\varepsilon,R_{c})=\varepsilon\chi(\varepsilon,R_{c})\, , \varepsilon=const. \label{auotovalori1}
\end{equation}

The function $\chi$ has to satisfy the boundary conditions $\chi(R_{c}=0)=0,\chi(R_{c}\rightarrow\infty)=0$. Now we take for $\chi$ the following expression 

\begin{equation}
\chi=\omega(\varepsilon,R_{c})exp\left\{-\frac{3\pi}{4l_{Pl}^{2}}(R_{c}-\frac{2\sigma^{2}}{\hbar})^{2}\right\} \label{soluzione}
\end{equation}

and $\omega$ is provided by the equation (here we adopt, for the kinetic term of the super-Hamiltonian, the same normal ordering as in \cite{Mon2002}, i.e. $p_{R_{c}}^{2}/R_{c}\rightarrow-\hbar^{2}\partial_{R_{c}}(1/R_{c})\partial_{R_{c}}$)

\begin{eqnarray}
\frac{l_{Pl}^{2}\hbar}{3\pi}\left\{\frac{\partial_{R_{c}}^{2}\omega}{R_{c}}-\partial_{R_{c}}\omega\left(4\alpha-
\frac{4\alpha R_{0}}{R_{c}}+\frac{1}{R_{c}^{2}}\right)\right.+\nonumber\\
+\left.\omega\left(4\alpha^{2}R_{c}-8\alpha^{2}R_{0}+\frac{4\alpha^{2}R_{0}^{2}}{R_{c}}-\frac{2\alpha R_{0}}{R_{c}^{2}}\right)\right\}+\nonumber\\
+\frac{\mu^{2}}{R_{c}}\;\omega-\frac{3\pi\hbar}{4l_{Pl}^2}R_{c}\omega+\frac{\sigma^{2}}{R_{c}^{2}}\omega-\frac{\epsilon}{c}\;\omega=0. \label{autoval2}
\end{eqnarray}

The solution of the above equation is the following Fuchs function

\begin{equation}
\omega=\sum_{n=0}^{\infty}c_{n}(\varepsilon)R_{c}^{n+2},
\end{equation}

whose coefficients obey to the following relations

\begin{align}
c_{n}=&-\frac{1}{n(n+2)}\left[(n+1)\frac{6\pi}{l_{Pl}^{2}}\sigma^{2}c_{n-1}\right.+\nonumber\\&+\left.\left(n\frac{3\pi}{l_{Pl}^{2}}+\frac{3\pi}{l_{Pl}^{2}\hbar}(\mu^{2}+\frac{3\pi}{l_{Pl}^{2}\hbar}\sigma^{4})\right)c_{n-2}\right] \label{fuchs}
\end{align}

Since $R_{c}\in[0,\,\infty)$, the series radius of convergence has to diverge and therefore we have

\begin{equation}
\lim_{n\rightarrow\infty}\frac{c_{n+1}}{c_{n}}=0;
\end{equation}

hence we see in such a limit the relation (\ref{fuchs}), in the leading order, rewrites

\begin{equation}
\frac{c_{n+2}}{c_{n}}\sim \frac{3\pi}{l_{Pl}^{2}}\frac{1}{n}.
\end{equation}

Comparing this behaviour to the corresponding one for the exponential term in (\ref{soluzione}), we conclude that the Fuchs series has to be truncated in correspondence to a given integer $m$ and the following relation holds

\begin{equation}
m+2=\frac{l_{Pl}^{2}\varepsilon^{2}}{3\pi\hbar^{2}c^{2}}+\frac{\mu^{2}}{\hbar}. \label{spettro}
\end{equation}

Equation (\ref{spettro}) comes out taking into account that the wave function (\ref{soluzione}) is solution of equation (\ref{autoval2}) if the key relation takes place

\begin{equation}
\varepsilon=-\frac{3\pi c}{l_{Pl}^{2}}\sigma^{2} \label{epsig}.
\end{equation}

Finally we stress how a notion of probability distribution can be associated to this solution, in agreement to the discussion presented in \cite{Mon2002}.

The relation (\ref{spettro}) provides a constraint on $\varepsilon$ and $\mu^{2}$ which allows that the spectrum of the total energy of the Universe is unbounded. However such a feature is removed as soon as we include in our dynamical scheme a cut-off on admissible physical lenghts; indeed there are clear indications (coming either from the Loop Quantum Gravity theory \cite{Rov1998}, either from the String approaches \cite{Polchinski}) that the space-time must have a ``lattice'' structure on the Planckian scale; in this respect we will require that every minimal lenght has to be equal to the Planck one $l_{PL}=\sqrt{G\hbar/c^{3}}$. Now we observe that the ultra-relativistic energy density can be written as follows:

\begin{equation}
\rho_{ur}=\frac{\mu^{2}c}{R_{c}^{4}}=\frac{3}{2}\frac{\hbar c}{d^{2}l_{Pl}^{2}},
\end{equation}

where $d$ denotes the cosmological horizon.\\
On the other hand the \emph{lenght per particle} of a perfect gas reads as

\begin{equation}
l^{3}\equiv\frac{V}{{\cal N}}=\frac{3}{2}\frac{\hbar^{2}}{m_{Pl}}\frac{1}{\rho_{gp}\lambda^{2}},
\end{equation}

being $m_{Pl}$ the Planck mass, $\lambda$ the thermal lenght of the particles and $\rho_{gp}$ the perfect gas energy density, i.e. $\rho_{gp}=\sigma^{2}c/R_{c}^{5}$. Thus if we require $d>l_{Pl}$ and $l>l_{Pl}$ we get the following inequalities for the quantum numbers $\sigma ^{2}$ and $\mu^{2}$ respectively

\begin{eqnarray}
\sigma^{2}\le\frac{3\hbar^{2}}{2m_{Pl}c}\left(\frac{R_{c}}{l_{Pl}}\right)^{5}\\
\mu^{2}\le\frac{3\hbar}{2}\left(\frac{R_{c}}{l_{Pl}}\right)^{4}.
\end{eqnarray}

Taking the minimal radius of curvature as that one at the Planckian time, i.e. 

\begin{equation}
R_{c}^{Pl}=\frac{R_{c}^{today}}{1+z^{Pl}}
\end{equation}

(being $z^{Pl}$ the Planckian redshift), then by (\ref{epsig}) we arrive to the constraint

\begin{equation}
\varepsilon\ge-\frac{9\pi}{2}\frac{\hbar^{2}}{l_{Pl}^{7}m_{Pl}}\left(\frac{R_{c}^{today}}{1+z^{Pl}}\right)^{5}.\label{stimaeps}
\end{equation}

Since $z^{Pl}$ must have a finite value we see how requiring a cut-off on the admissible lenghts, yields a lower bounded spectrum.

\section{Inhomogeneous Spherical Perturbations}
\label{sec:InhomogeneusSphericalPerturbations}

In order to make account for the quantum dinamics of inhomogeneous perturbations we consider the spherically simmetric Lemaitre-Tolman \cite{Peebles,Lem,Tol} model, whose spatial line element reads

\begin{equation}
dl^{2}=e^{2\alpha}dr^{2}+e^{2\beta}(d\theta^{2}+sin^{2}\theta d\phi^{2}),
\end{equation}

where the functions $\alpha(t,r), \beta(t,r)$ take the explicit form

\begin{equation}
e^{\alpha}=\frac{(a(r,t)r)'}{\sqrt{1-r^{2}/R^{2}}}, \quad e^{\beta}=ra(r,t)
\end{equation}

being $R$ a given constant and with $(\quad)'\equiv d(\quad)/dr$. In order to study inhomogeneous perturbations \cite{Kolb} we write the function $a(t,r)$ as follows

\begin{equation}
a(t,r)=R_{c}(t)+\xi(t,r)
\end{equation} 

with $\mid\xi\mid\ll\,\mid R_{c}\mid$. Expanding the Einstein-Hilbert-matter action up to first order in $\xi$ we get in a sinchronous reference an action of the form (the spherical simmetry prevents to have a non zero shift vector)

\begin{equation}
S=\int dt \left\{p_{R_{c}}\frac{\partial R_{c}}{\partial t}+\int_{0}^{R}dr p_{\xi}\frac{\partial\xi}{\partial t}-(H_{0}+H_{1})\right\}, \label{azpert}
\end{equation}

where $p_{R_{c}}$ and $p_{\xi}$ denote tha coniugate momenta respectively to $R_{c}$ and $\xi$, $H_{0}$ is the FRW Hamiltonian and $H_{1}$ reads

\begin{equation}
H_{1}=H_{1}^{\xi}+H_{1}^{M}
\end{equation}

with

\begin{align}
H_{1}^{\xi}&=\int_{0}^{R}dr A(R_{c},r)\xi\\
H_{1}^{M}&=\int_{0}^{R}dr [B(R_{c},r)\delta\mu^{2}(r)+C(R_{c},r)\delta\sigma^{2}(r)];
\end{align}

the above functions $A, B, C$ correspond to the expressions

\begin{align}
A(R_{c},r)&=-\frac{2G}{(3\pi)^{2}Nc}\frac{P_{R_{c}}^{2}}{R_{c}^{2}}(F'_{3}-F_{2})-\frac{Nc^{3}}{2G}M_{0}+\nonumber\\&+4\pi^{3/2}N\frac{\sigma^{2}}{R_{c}^{3}}\Sigma+3\sqrt{\pi}N\frac{\mu^{2}}{R_{c}^{2}}\Sigma,\\
&B(R_{c},r)=3\sqrt{\pi}N\frac{F_{2}}{R_{c}},\\ &C(R_{c},r)=2\pi^{3/2} N\frac{F_{2}}{R_{c}^{2}}.
\end{align}

where we used the definitions

\begin{equation}
F_{n}(r)=\frac{r^{n}}{\sqrt{1-r^{2}/R(r)^{2}}},\quad \Sigma(r)=\frac{(F_{2}r)'}{3}-F_{2}.
\end{equation}

$H_{1}^{\xi}$ corresponds to the pure geometrical contribution to the perturbed Hamiltonian, while $H_{1}^{M}$ makes account for the corresponding perturbation in the matter fields (both the ultra-relativistic matter and the perfect gas are taken, in the leading order, comoving to the expansion).

Now we observe that from the action (\ref{azpert}) it follows $\partial_{t}\xi=0$, i.e. $\xi=\xi(r)$; such a behaviour allows to identify $\xi$ with the matter fields perturbations and, by comparing the same inverse powers of $R_{c}$ appearing in $H_{1}^{\xi}$ and $H_{1}^{M}$, we get the key relation

\begin{equation}
\xi(r)=-\frac{2}{3}\pi\frac{F_{2}}{\Sigma}\frac{\delta\sigma^{2}(r)}{\mu^{2}}.
\end{equation}

The eigenvalue problem (\ref{autovalori0}) takes, for such perturbed FRW model, the smeared form

\begin{eqnarray}
\left(\hat{H}_{0}+\hat{H}_{1}\right)\psi(R_{c},r)=\nonumber\\
=\left[\varepsilon +\int_{0}^{R}4\pi r^{2} \Delta\rho(r)dr\right]\psi(R_{c},r).
\end{eqnarray}

From the time independent perturbation theory \cite{Sakurai} on $\hat{H_{1}}$ (which now depends only on $\delta\sigma^{2}$ and $\delta\mu^{2}$), we find for $\Delta\rho(r)$ the following expression

\begin{equation}
\Delta\rho(r)=D(r)\delta\sigma^{2}+E(r)\delta\mu^{2}\label{denspert}
\end{equation}

being

\begin{align}
D(r)=&-\frac{G}{3c^{2}} \left(F'_{3}-F_{2}\right)\frac{F_{0}}{\Sigma}\frac{1}{\mu^{2}}\left[\frac{\hbar}{3\pi l_{Pl}^{2}}\mu^{2}I_{2}-\frac{\hbar}{2l_{Pl}^{2}}I_{0}\right.+\nonumber\\&+\left.\frac{\hbar}{3\pi l_{Pl}^{2}}\sigma^{2}I_{3}-\frac{\hbar}{3\pi l_{Pl}^{2}}\frac{\epsilon}{c}I_{1}\right]
-\frac{2}{3}\frac{\pi^{3/2}}{c}\frac{\sigma^{2}}{\mu^{2}}I_{3}F_{0}+\nonumber\\&+\frac{c}{12G} I_{0}\frac{1}{\mu^{2}}\frac{1}{\Sigma}\label{D}\\ 
E(r)=&\frac{3}{4}\frac{cI_{1}}{\sqrt{\pi}}F_{0}. \label{E} 
\end{align}

Above by $I_{k}, k=1,2,3....$ we denoted the corresponding  mean values

\begin{equation}
I_{k}=\int_{0}^{\infty} dR_{c}\frac{\chi_{0}^{\ast}\chi_{0}}{R_{c}^{k}}
\end{equation}

we stress that that $I_{1}, I_{2}, I_{3}$ appearing in (\ref{D}) and (\ref{E}) correspond to finite values because the first contribution in $\omega(R_{c})$ is of order $R_{c}^{2}$.

\section{Phenomenological implications}
\label{sec:PhenomenologicalImplications}

As shown in \cite{CosmIssues} in the classical limit, as taken in the WKB approach, the critical parameter $\bar{\Omega}_{dm}$ for our dark matter candidate reads in term of $\varepsilon$ (which behaves like a constant of the motion) in the form

\begin{equation}
\bar{\Omega}_{dm} = -\frac{4l_{Pl}^{2}c\varepsilon}{3\pi\hbar H^{2}(R_{c}^{today})^{3}}. \label{odm}%
\end{equation}

The value of $\varepsilon$ which ensures that $\bar{\Omega}_{dm}$ is today of order unity is estimated to be $\varepsilon\sim\mathcal{O}(-10^{82})GeV$. Now if we take $\varepsilon$ from the inequality (\ref{stimaeps}), we arrive to the final inequality for $\bar{\Omega}_{dm}$

\begin{equation}
\bar{\Omega}_{dm}\ge\frac{6c\hbar}{H^{2}m_{Pl}}\left(\frac{1}{l_{Pl}(1+z^{Pl})}\right)^{5}(R_{c}^{today})^{2}.\label{stimaodm}
\end{equation}

In agreement to the standard interpretation of the quantum mechanics we expect that the Universe settled down into the state of minimal energy and this is equivalent to take in equations (\ref{stimaeps}) and (\ref{stimaodm}) the equality sign. Being $R_{c}^{today}$ estimated of the order $\mathcal{O}(10^{28})cm$ \cite{Map} then we get $\bar{\Omega}_{dm}\sim\mathcal{O}(1)$ as soon as $z^{Pl}\sim\mathcal{O}(10^{50})$; such a value of the Planckian redshift is compatible with a smooth inflationary scenario whose \emph{e-folding} can be estimated of the order $\sim 50$.\\ We remark that the case in which the Universe did not undergo an inflationary scenario, would correspond to $z^{Pl}\sim\mathcal{O}(10^{30})$; such a value of $z^{Pl}$ fixes the lower boundary for $\varepsilon$ in (\ref{stimaeps}) like $\mathcal{O}(-10^{139})GeV$.

In such a case the value of $\varepsilon$ which provides today $\bar{\Omega}_{dm}\sim\mathcal{O}(1)$ is no longer the minimum of the spectrum but it lies within it. The case of the Standard Cosmological Model (SCM) ($z^{Pl}\sim\mathcal{O}(10^{30})$ is of particular interest here because if we take the value of $\sigma^{2}$ corresponding to $\varepsilon\sim\mathcal{O}(-10^{82})GeV$ and evaluate $\mu^{2}$ via its maximum value (for such values in the spectrum relation (\ref{spettro}) the $\varepsilon$-term dominates the $\mu^{2}$-one), then the perturbation density (\ref{denspert}) rewrites, in the leading order, as

\begin{equation}
\Delta\rho=\frac{3}{4}\frac{cI_{1}}{\sqrt{\pi}}F_{0}\delta\mu^{2}.\label{rhomu}
\end{equation}

The above expression acquires particular interest because it outlines the expected direct correlation between the perturbations in our dark matter candidate and the ultra-relativistic matter. Summarizing we constructed an evolutive quantum cosmology which leads to a non-zero super-Hamiltonian eigenvalue and outlined how it can be a satisfactory dark matter candidate. In particular the perturbation theory of the eigenvalue problem allowed us to fix the dark matter perturbation in terms of those ones in the ultra-relativistic matter and in the perfect gas, with the relevant issue (\ref{rhomu}) for the SCM case. Our result calls attention for a deep investigation of the semiclassical limit which allows to precise the mechanism by which the quantum perturbation (\ref{denspert}) are frozen out and approach a classical limit.

\end{document}